\documentclass[12pt]{elsart}

\begin{document}

\def\beq{\begin{eqnarray}}
\def\eeq{\end{eqnarray}}

\mbox{} \hfill ISGBG-02 \\
\begin{frontmatter}

\title{\sc Wave-Particle duality at the Planck scale:
Freezing of neutrino oscillations}

\author{D. V. Ahluwalia}

\address{ISGBG, Ap. Pos. C-600, Escuela de Fisica de la UAZ\\
Zacatecas 98068, Mexico; E-mail: ahluwalia@phases.reduaz.mx, \\
Fax: +52 492 44 55 2, http://phases.reduaz.mx/}

\begin{abstract}
A  gravitationally-induced modification to de Broglie
wave-particle duality is presented.
At Planck scale, the gravitationally-modified 
matter wavelength saturates to a few times the 
Planck length in a momentum independent manner. In certain frameworks, 
this circumstance freezes neutrino oscillations
in the Planck realm. This effect is apart, and beyond, 
the gravitational red-shift.
A conclusion is drawn that in a complete theory of quantum
gravity the notions of  ``quantum'' and ``gravity''  shall carry
new meanings -- 
meanings, that are yet to be deciphered from theory and observations 
in their entirety.
\end{abstract}

\end{frontmatter}

\section{Introduction}

One of the most challenging quests in contemporary 
theoretical physics concerns
the nature of space-time at the Planck scale, and  
deciphering gravi\-tatio\-nal\-ly-induced modifications to the quantum 
realm, and vice versa. While some of these aspects can only be revealed
by observations, Maxwellian arguments of consistency
can also shed light on the joint realm of the gravity and the 
quantum. 

One such  Maxwellian argument was presented in Ref.
\cite{dva_grf94}. It says that quantum 
measurements in the Planck realm necessarily
alter the local space-time metric in a manner that 
destroys the commutativity of the position measurements of two different
particles. In addition, it also
affects the fundamental commutator,
\beq
\left[x,p_x\right]=i\hbar\label{fc}
\eeq
The essential idea of the above argument resides in the observations that a 
position measurement collapses the wave function, say, in the following 
manner,
$\mbox{Position Measurement}: \langle\vec r\,
\vert\psi(0\le r\le \infty)\rangle \rightarrow
\langle \vec r\,\vert\psi(0\le r\le R)\rangle
$.
In case $R$ is of the order of Planck length, the gravitational effects
associated with the wave function collapse become important as it
necessarily invokes the collapse
of the energy-momentum tensor. Hence, the  local space-time metric changes.  
As shown in Ref. \cite{dva_grf94}, this circumstance 
makes the position measurements of two 
distinguishable particles non-commutative.

As a consequence,  non-locality must be an essential part of any attempt 
to merge the theory of general relativity with quantum mechanics. The 
derived non-commutativity easily extends to measurements of different 
components of the position
vector of a single particle, and modifies the fundamental commutators of the
Heisenberg algebra. 
A further essential conclusion beyond the stated 
gravitationally-induced non-locality is that space-time itself 
acquires a non-commutative character.

Some implications of such 
non-com\-mut\-ati\-ve space times have been studied, e.g., by Madore 
\cite{jm}, and by Connes \cite{ac}, however, from an
entirely different view point.  
Independently, efforts in string theories  
also arrive at gravitationally-modified fundamental 
uncertainty relations. In that context 
an early reference is the work of Veneziano \cite{v}, while a recent 
one is \cite{sdh}. In yet another line of argument,
without invoking extended objects, and entirely within the 
framework of quantum mechanics and the theory of general relativity,
Adler and Santiago \cite{as} also obtain similar
modifications to the uncertainty principle without 
invoking extended objects ({\em cf.\/},  
\cite{gac,mm,ns,fs}).
A somewhat different argument, based on the existence of an upper bound
for acceleration, also results in a gravitational modification to the
uncertainty principle \cite{cls}.
The mathematical expression of the above results that leads to a 
gravitationally
modified expression for the wave particle duality is given by
the following modification to the fundamental commutator \cite{ak}:
\beq
\left[{\bf x},\,{\bf p}\right] =i\hbar \left[{\bf 1} +\epsilon \frac{ 
\lambda^2_P {\mathbf p}^2}
{\hbar^2} \right]\label{fun_comm}
\eeq
where $\lambda_{P} = \sqrt{\hbar G/c^3}$, is the Planck length, 
and $\epsilon$
is some dimensionless number of the order of unity. In what follows
I set $\epsilon$ equal to unity.

It is the purpose of this {\em Letter\/} to 
decipher the wave-particle duality as contained in  (\ref{fun_comm}).  
To make our argument, we first recapture the origin of the wave-particle
duality in the absence of gravitational effects, and then immediately 
return to the stated objective.

\subsection{Wave-Particle duality in the absence of gravity}

The fundamental commutator, (\ref{fc}), 
encodes the fact that intensity of matter and gauge fields cannot be 
arbitrarily reduced to zero, but is bounded from below.
The first direct evidence for this circumstance came from Einstein's 
understanding of the photo-electric effect.
It is precisely this commutator that lies behind the de Broglie 
relation, and the entire edifice of the wave-particle duality. To see this,
recall that in configuration space,  
$p_x=\frac{\hbar}{i}\frac{\partial}{\partial x}$,
is a solution of the fundamental commutator (\ref{fc}), with eigenfunctions
of the form $\psi(p_x) = N \exp\left(\frac{i}{\hbar} p_x x\right)$. The
spatial periodicity, $\lambda = \frac{h}
{\vert p_x\vert}$, carried by 
 $\psi(p_x)$, when extended to three dimensions, yields 
the well known de Broglie relation 
\beq
\lambda = \frac{h}{p},\label{db}
\eeq
where $p=
\vert \vec p\,\vert$ is the magnitude of the momentum vector
associated with an object. 
A simple text-book algebraic exercise, with (\ref{fc}) as the
physical input, gives the Heisenberg uncertainty relation
\beq
\Delta x\Delta p_x \ge \hbar/2,\,\,\mbox{etc.}\label{hur}
\eeq

In the absence of gravity,
equations (\ref{fc}), (\ref{db}), and (\ref{hur}) represent
various inter-related aspects
of the  wave-particle duality. One immediately sees
that as $p$ approaches the Planck scale, and then beyond, the de Broglie
wavelength continuously shrinks to zero and allows quantum-mechanical
probing of space-time to all length scales and energies. 
However, as already mentioned,  if the gravitational
effects in the quantum-measurement process are taken into consideration, 
these results are no longer true. Planck length, up to a factor of the order
of unity, emerges as the 
limiting length scale beyond which space-time cannot be probed.
This circumstance, therefore, immediately suggests
that the the relation (\ref{db}) must undergo a change in which 
the left hand  side of (\ref{db}) saturates to, within a few times, the  
Planck length. It is precisely this that emerges in the following.

As long as the entire theoretical structure of the existing quantum field 
theories rests upon the wave-particle duality,
it is necessary to fix the domain of its validity. 
The heaviest objects for which the wave-particle duality (\ref{db})
has been experimentally verified, so far, is
the $C_{60}$ fullerene \cite{f}. In this context, the experiment 
sets the scale at $m_{C_{60}}= 1.20\times 10^{-21}$ g. 
This is already an impressive achievement. 
Yet, it is to be compared with the Planck mass,
$m_{Pl.}\equiv(\hbar c/G)^{1/2} = 2.18\times 10^{-5}$ g.  
However, in order to study  possible 
departures from the de Broglie 
wave-particle duality in the Planck regime, 
one may even not need to invoke early universe directly.
All one may need are Planck mass quantum objects, 
and an appropriate technique 
to study an associated interference phenomena. To be specific, 
such effects may become
indirectly observable via extremely high-energy gamma rays, and 
high-energy neutrinos.

Here, and in the following, the
operator, or $c$-number, nature of objects, such as $p_x$ in (\ref{fc}), 
where it is an operator, and $p_x$ in $\psi(p_x)$, where it is a $c$-number,
shall be omitted and will be assumed apparent from the context.

\section{Gravitationally-modified Wave-Particle duality: 
minimal modification, and some implications}

In one spatial dimension (chosen as $x$), 
the gravitationally-modified position-momentum uncertainty relation
immediately follows from the commutator (\ref{fun_comm}), and reads: 
\beq
\Delta x\,\Delta p_x\ge \frac{\hbar}{2}\left[1+ \left(
\frac{\lambda_{P} \Delta p_x}{\hbar}\right)^2+ \left(
\frac{\lambda_{P} \langle {\mathbf p}\rangle}{\hbar}\right)^2\right].
\label{gmod_ur}
\eeq
It carries as a characteristic feature
the Kempf-Mangano-Mann (KMM, ref.\cite{ak})
lower bound on the position uncertainty:
\beq
\Delta x_{_K}=\lambda_{P}\left(1+
\frac{\lambda_{P} \langle {\bf p}\rangle}{\hbar}\right)^{1/2}
\eeq
Notice that $\Delta x_{K}$ has a state dependence via $\langle {\mathbf p} 
\rangle$. For a state of a vanishing $\langle {\mathbf p}\rangle$, one 
obtains 
the absolute minimal distance that can be probed quantum mechanically. 
This lowest bound does not depend on the particle species. Therefore,  
the existence of the ``absolute minimal distance'' 
suggests a new intrinsic property of the space-time itself.


An important implication of the KMM lower bound, $\Delta x_{_K}$, is that
the de Broglie plane waves can no longer represent the physical wave 
functions, even not in principle. 
Thus the wave-particle duality
must undergo a fundamental conceptual and quantitative change.

A non-relativistic modification to the de Broglie relation
was presented in  pioneering KMM work. 
This case, however, is likely to 
be of limited interest in the Planck regime. Here, I present the 
gravitationally modified de Broglie relation without restrictions 
on the particle's momentum. 

It is readily seen 
that the momentum space wave function consistent with 
the gravitationally modified uncertainty  relations (\ref{gmod_ur}) 
reads \cite{ak}:

\vbox{
\beq
\psi(p)&=&
N\left(1+\beta p^2\right)^{-\left[
\,{\kappa({\mathbf p})}/{4\beta(\Delta p)^2}\right]}\nonumber\\
&& \times\exp\left[
-i\,\frac{\langle{ {\mathbf x}\rangle}}{\lambda_{{P}}}
\tan^{-1}\left(\sqrt{\beta} p\right) 
- \frac{\kappa({\mathbf p}) \langle
{\mathbf p}\rangle}{2 (\Delta p)^2\sqrt{\beta}}
\tan^{-1} \left(\sqrt{\beta} p\right)\right]
\eeq
}
where $\kappa({\mathbf p}):=1+\beta(\Delta p)^2+\beta\langle
{\mathbf p}\rangle^2$, and $\beta:=\lambda^2_P/\hbar^2$. $N$ is a 
normalization factor.
This represents an oscillatory function damped by a momentum-dependent
exponential. I identify the oscillation length with the gravitationally
modified de Broglie wave length:
\beq
\lambda= 2\pi\,\frac{\lambda_{{P}}}{\tan^{-1} \left(\sqrt{\beta} p\right)}
\eeq
Introducing $\overline{\lambda}_{{P}}:={2\pi \lambda_{{P}}}$ as the
{\em Planck circumference}; 
and $\lambda_{{dB}}$ as the gravitationally {\bf un}modified de Broglie
wave length,  $\lambda_{{dB}}=h/p$, the above expression takes
the form:
\beq
\lambda=
\frac{ {\overline{\lambda}_{{P}}} }
{\tan^{-1}\left(\overline{\lambda}_P /\lambda_{{dB}}\right)}
\cases{\rightarrow\lambda_{dB} & for low energy regime  \cr
\rightarrow 4\lambda_{{P}} &
for Planck regime  \cr}\label{lambda}
\eeq
In addition, for the specific non-relativistic regime considered 
by Kempf {\em et al.} \cite{ak},
$\lambda$ reproduces their equation (44). This justifies the interpretation
of the oscillatory length associated with KMM's $\psi(p)$ as the
gravitationally modified de Broglie wavelength.

The gravitationally induced modifications to
(\ref{fc}), (\ref{db}), and (\ref{hur}) are now contained in
(\ref{fun_comm}), (\ref{lambda}), and (\ref{gmod_ur}). These
latter equations constitute the minimal  conceptual and quantitative changes
in the nature of the wave-particle duality.

A brief discussion on immediate physical implications of the
modified wave-particle duality in the Planck realm now follows. 

\subsection{Freezing of neutrino oscillations}

To explore one of the concrete consequences of the above-presented
modification to the
wave-particle duality,  note that the
existing data suggests flavor eigenstates of neutrinos to be
linear superposition of different mass eigenstates \cite{data}:
\beq
\vert
\nu_\ell\rangle = \sum_\jmath U_{\ell\jmath} \vert 
m_\jmath \rangle\label{def}
\eeq
where, $\ell=\e,\mu,\tau$, is the flavor index, and  $\jmath=1,2,3$, 
enumerates the mass eigenstates, while 
$U$ is a $3\times 3$ unitary matrix. Several 
fundamental questions now arise.
Is this low-energy, i.e. low in comparison to the Planck mass, 
construct still valid at the Planck scale?
What is the time evolution of the flavor and mass eigenstates
in the Planck realm?
At a deeper and related level, does the non-commutative space-time 
still carry Poincar\'e symmetry? -- for the very notions of mass and 
spin (which the underlying mass eigenstates carry) originate from
the Casimir invariants associated with the Poincar\'e group. In addition,
the equations governing the evolution of the states derive their form
from the space-time symmetries. None of these questions has a readily 
available answer. An answer must, therefore, await future theoretical and 
observational input. The latter, for example,  may come from the study of 
anomalous events around and beyond $10^{20}$ eV
cosmic rays.

Under these circumstances we take note of the fact that low-energy neutrino
oscillations owe their physical origin to different de Broglie
oscillation lengths associated with each of the underlying mass eigenstates.
If one assumes that each of the mass eigenstates carries the same energy, then
the flavor oscillations arise due to different 
de Broglie oscillation lengths carried by each of the mass eigenstates. 
If this scenario was considered for neutrino oscillations then it is 
clear that neutrino oscillations shall freeze at Planck scale
due to the above obtained
gravitationally-induced modification to 
the wave-particle duality. In particular, I draw attention to the
saturation of $\lambda$ as indicated in Eq. (\ref{lambda}).

In the ordinary neutrino oscillation phenomenology the flavor oscillations
are not altered at any practical level if one considers the ``equal energy,''
or ``equal velocity'', 
or ``wave packet'' approaches \cite{da1,da2,da3,da4}. Not
knowing the answer to the questions posed above it is not yet possible to say
if the Planck-scale freezing of neutrino flavor oscillations shall survive
in all neutrino oscillation frameworks. 

Our discussion here, therefore, is intended to bring attention to the fact that
the gravitationally-induced modifications to the wave-particle duality
may have significant physically observable consequences for the early 
universe. 

\subsection{Effect on H-atom}

For comparison, 
to the lowest order in $\lambda_{P}$, the effect of the modification 
(\ref{gmod_ur}) on the ground state level of the hydrogen atom results 
in the following modified uncertainty principle estimate for the 
ground state of an electron in an H-atom:
\beq
\left(E_0\right)_g \simeq - \,\frac{m e^4}{2\hbar^2}
\left[
1- \frac{4 m \lambda^2_P}{\hbar^2}\left(\frac{m e^4}{2\hbar^2}\right)
\right]
\eeq
Identifying:
\beq
E_0 = - \,\frac{m e^4}{2\hbar^2}
\eeq
with the ground state level of the hydrogen atom without
incorporating the gravitationally-induced correction to the
uncertainty relation, 
one immediately notices that
the effect of gravitational corrections   
is to reduce the magnitude of the 
ionization energy by $2.5\times 10^{-48}\,\,\mbox{eV}$. 
This suggests that a space-time endowed with the KMM bound is in some
sense a heat bath as it 
decreases the energy required to disassociate 
the H-atom.

\subsection{Coherence in the early universe and in biological systems}

The wavelength $\lambda$ asymptotically approaches
the constant value $4\lambda_{{P}}$ that is now universal 
for all particle species. As a consequence 
of this universality, a new type of coherence may emerge 
in the early universe and this may carry 
significance for the large-scale uniformity
of the universe. It is also speculated that quantum mechanics
plays a fundamental role in brain function, see,e.g., \cite{qb}. Therefore, the new 
coherence may also
carry significant implications for functioning of the brain, and
other biological systems, if important biological elements carry
a mass of the order of $M_P=\left(\hbar c/G\right)^{1/2} = 2.2\times
10^{-5}\,\,\mbox{g}$.

\section{Conclusion}

If the effects of the gravitationally-induced modification to the
de Broglie wave particle duality are  
negligible for low energy, their relevance perhaps cannot
be overestimated at the Planck-scale. At present, there are 
already speculations that anomalous events around $10^{20}$ eV
cosmic rays may be pointing towards a violation of the Lorentz symmetry 
\cite{lgm,cg}.
It is expected that the gravitationally-modified wave-particle duality
carries with it deformations of the Poincar\'e symmetries.
Some of these deformations can be studied with the recently-approved 
Gamma-ray Large Area Space Telescope (GLAST), and with other
detectors.\footnote{The reader is referred 
to references  \cite{p4,grb,gac_tp} 
for the original proposal, and for details on the
recent progress in this direction. A related proposal on 
gravitationally-induced modification of quantum evolution
by Ellis, Hagelin,
Nanopoulos, and Srednicki \cite{ehns} can be studied via flux equalization
of the cosmic neutrinos as shown by Liu, Hu, and Ge \cite{yong}.
The possible freezing of neutrino oscillations in the early universe
could carry significant impact on the formation of structure in the 
early universe.}

In the context of this {\em Letter\/}, and two recent works 
\cite{qVEP,dva_grf2000},\footnote{Under the assumption
of operationally independence of the inertial and
gravitational masses, these works establish
a quantum-induced violation of the equivalence principle.}
the above discussion makes it clear that the conceptual foundations 
of the  theory of general relativity and quantum mechanics 
are so rich that they impose concrete modifications 
onto each other in the interface region. Yet, a complete theory of quantum
gravity shall carry ``quantum'' and ``gravity'' with new meanings - meanings
that are yet to be deciphered from theory and observations in their entirety.

\section*{Acknowledgements.} 
It is my pleasure to thank Achim Kempf for an extended
e-discussion on the subject, and Mariana Kirchbach for a 
critical reading of the manuscript. 
This work was supported by CONACyT (Mexico), 
and ISGBG.



\end{document}